\documentclass[twocolumn,prl,showpacs,superscriptaddress, nofootinbib]{revtex4}
\usepackage{amssymb}
\usepackage{amsmath}
\usepackage{graphicx}
\usepackage{dcolumn}
\usepackage{bm}
\usepackage{epsfig}
\usepackage{color}
\usepackage[T2A]{fontenc}
\usepackage{enumitem}

\begin{document}

\title{Exciton radiative decay through plasmon modes in planar
metal-semiconductor structures}
\author{M. V. Durnev}
\affiliation{SOLAB, V. A. Fock Institute of Physics Saint-Petersburg State University,
198504 St.-Petersburg, Russia}
\affiliation{Ioffe Physical-Technical Institute of the RAS, 194021 St.-Petersburg, Russia}
\author{A. V. Kavokin}
\affiliation{SOLAB, V. A. Fock Institute of Physics Saint-Petersburg State University,
198504 St.-Petersburg, Russia}
\affiliation{Physics and Astronomy School, University of Southampton, Highfield,
Southampton, SO171BJ, UK; CNRS-Universit\'{e} Montpellier, Laboratoire
Charles Coulomb UMR 5221, F-34095 Montpellier, France}
\author{B. Gil}
\affiliation{CNRS-Universit\'e Montpellier, Laboratoire Charles Coulomb UMR 5221, F-34095
Montpellier, France}

\begin{abstract}
We develop a non-local dielectric response theory to describe the
temperature dependence of exciton lifetime in metal-semiconductor
heterostructures. Coupling between excitons and surface plasmons results in
a strongly nonmonotonous behaviour of exciton radiative decay rate versus
temperature. Tuning the plasmon frequency one can control the efficiency of
exciton emission of light.
\end{abstract}

\maketitle

Emission enhancement by placing a light source near the metal surface
generally known as the Purcell effect was studied for a wide variety of
metal-emitter structures~\cite{konderink, meinzer,
PhysRevLett.96.113002,PhysRevLett.97.017402,science_curto,Ajiki, glazov_ftt}%
. However in the spectral range where the dielectric function of metal
exhibits features linked to collective oscillations of electron gas one
should take into account the possible energy transfer from excitons to
plasmon modes. In this Letter we show that exciton-plasmon coupling may
dramatically affect the exciton radiative lifetime and strongly modify the
Purcell effect.

Energies of bulk plasmons in typical metals lie in the ultraviolet spectral
range ($\sim 10$~eV). Introduction of a boundary between metal and
dielectric medium results in appearance of a new type of coupled
light-electronic modes: surface plasmon-polaritons (SPP) whose resonance
frequencies may be found in the near ultraviolet or even optical spectral
region. Tuning the SPP frequencies to the vicinity of the band-gap frequency
in metal-semiconductor structures allows for coupling of semiconductor
emission to the plasmonic modes resulting in a strong modification of the
emitter characteristics. Firstly observed for emitting molecules placed near
the metal surface~\cite{Ford1984, Chance1978} this effect is now on topic
for the use in solid state physics, especially in gallium-nitride-based
light emitting devices, where increasing the light extraction efficiency is
one of the major technological challenges. The reduction of exciton lifetime
in GaN/InGaN emitters covered by metal layers has been observed
experimentally at the beginning of this century~\cite{yablonovitch1999,
yablonovitch2002, nat_mat_okamoto, okamoto:071102, lin:221104}. A similar
effect has been detected recently in the structures containing metallic
nanoparticles~\cite{Jang:12, toropovPRL2009} and cells~\cite{TanakaPRL2010}.

\begin{figure}[hptb]
\includegraphics[width=0.3\textwidth]{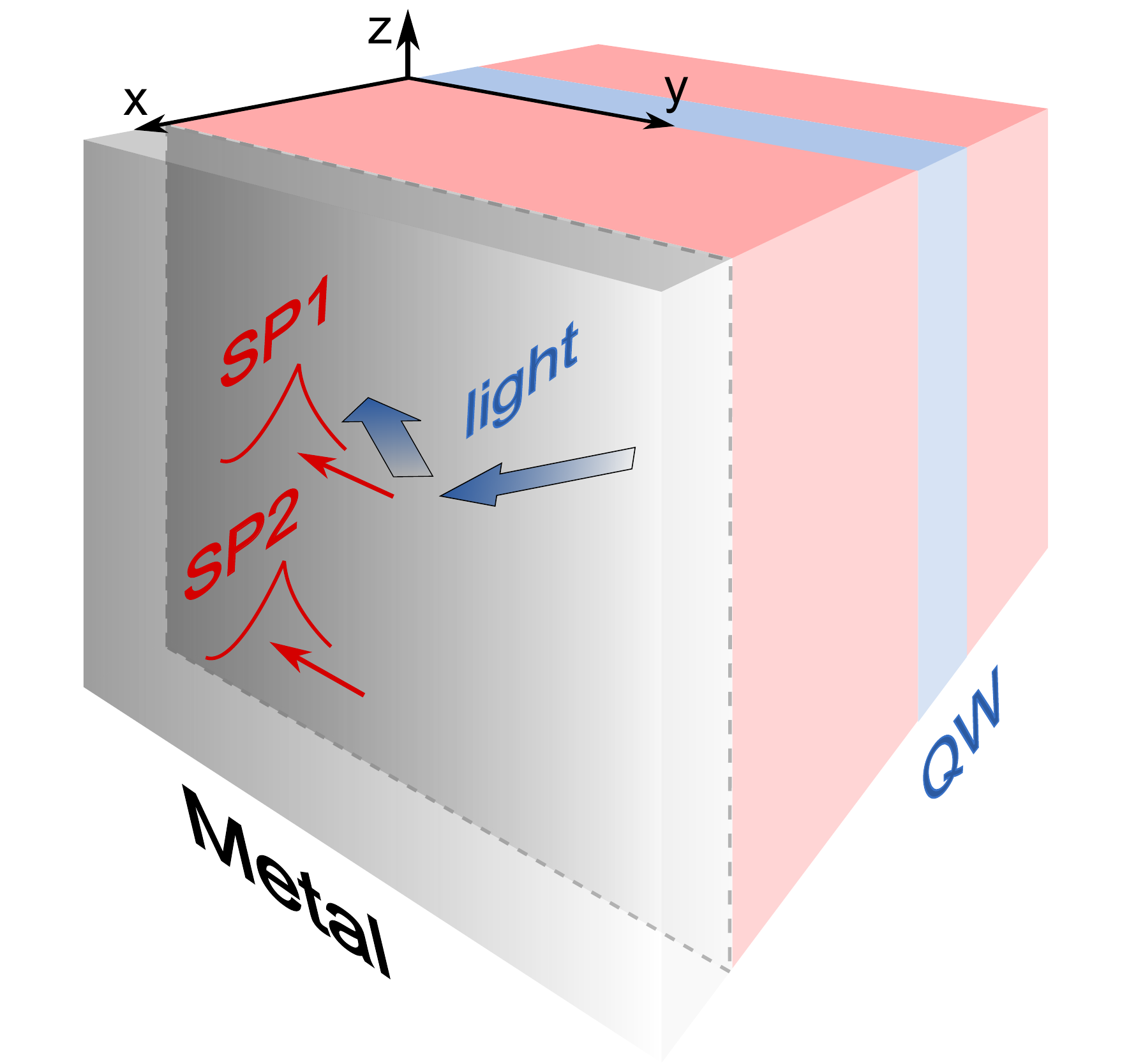} \vskip0.5cm %
\includegraphics[width=0.3\textwidth]{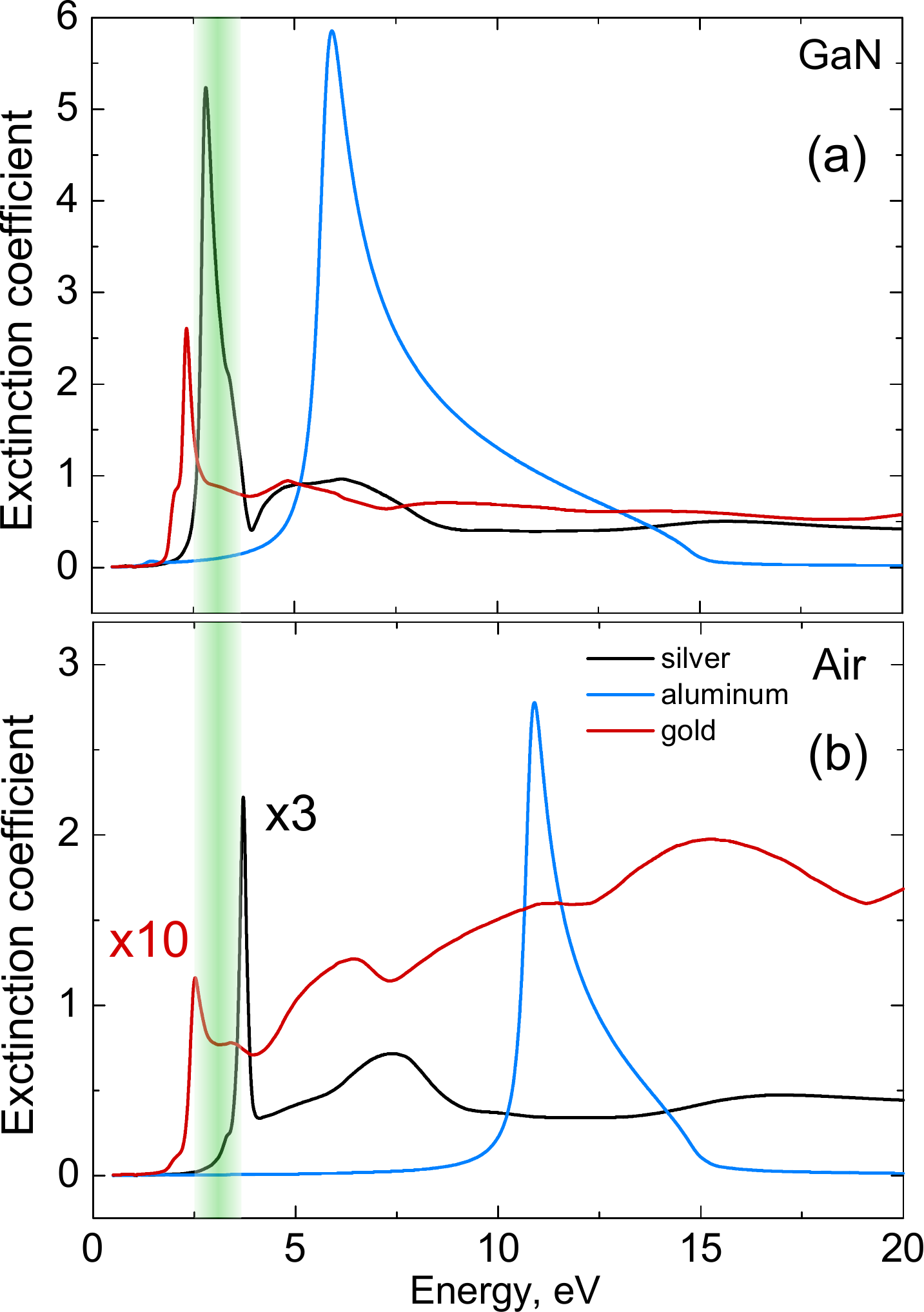}
\caption{(Upper panel) The schematic illustration of the structure, (lower
panel) Imaginary parts of effective dielectric function~(\protect\ref%
{eq:dispersion}) for different metals and values of $\protect\varepsilon_D$:
(a) $\protect\varepsilon_D = 6.1$ corresponding to GaN and (b) $\protect%
\varepsilon_D = 1$. The green area presents the typical range of exciton
energies in InGaN quantum well.}
\label{fig:fig1}
\end{figure}

Here by means of a non-local dielectric response theory we describe the
temperature dependence of exciton lifetime in planar metal-semiconductor
heterostructures. We show that the exciton radiative decay affected by the
coupling with surface plasmon modes, experiences a strong nonmonotonous
variation with temperature. Tuning the plasmon frequency as well as the
exciton level position in a quantum well (either by quantum well width or
temperature variation) one can achieve a significant enhancement of the
exciton emission efficiency. The structure considered in this paper is
schematically shown in Fig.~\ref{fig:fig1}. It consists of a quantum well
(QW) sandwiched between two barrier layers, one of them covered by a
metallic film. In the calculations we use the parameters of InGaN/GaN QWs,
which are mostly experimentally relevant, although our model is applicable
for a wide variety of materials.

Let us first consider free excitons in a quantum well with a broad thermal
distribution of energies (this model is well applicable for light emitting
diode devices, where the charge carriers are injected in an active region
electrically). The excitons therefore have in-plane wavevectors $%
k_{\parallel }$ varying in large limits with the corresponding occupation
numbers obeying the Boltzmann statistic. Optical selection rules require
conservation of the in-plane wave-vector for light emitted by excitons in
QWs. Light emitted by excitons propagates inside the barrier layers and
either escapes the system through its boundary or excites one of the plasmon
mode or, after a reflection act, comes back to QW and eventually decays in
the substrate material. It is well known that only the photons with
wavevectors inside the so-called light cone can escape the ideal planar
sample. Thus only the excitons with the wavevectors satisfying the condition
$k_{\parallel }<\omega /c$, where $\omega $ is the light frequency and $c$
is the speed of light in vacuum are directly coupled to the continuum of
photonic states in vacuum. The decay rate of these excitons depends on the
polarisation of light \cite{ivchenko05a}. 
The heavy-hole excitons in zinc-blende-based semiconductor heterostructures
as well as A-excitons in GaN/InGaN systems are optically inactive in
polarization $\bm{e}\parallel x$~\cite{ivchenko05a}, where $x$ is the growth
direction, so that the corresponding contribution in \textit{p}-polarized
emission vanishes. 

The integrated decay rate of the thermal population of excitons in a quantum
well can be obtained by simply averaging single exciton decays over all
values of $k_{\parallel }<\omega /c$. Considering isotropic isoenergetic
contours of an exciton in $k_{\parallel }$-space and a Boltzmann statistic
for the excitons at temperature $T$ one can obtain 
\begin{equation}
\Gamma (T)=\frac{\hbar ^{2}}{Mk_{B}T}\sum\limits_{i=s,p}\int\limits_{0}^{%
\omega /c}k_{\parallel }dk_{\parallel }\exp \left( -\frac{\hbar
^{2}k_{\parallel }^{2}}{2Mk_{B}T}\right) \Gamma _{0,i}(k_{\parallel }):,
\label{eq:average}
\end{equation}%
where $M$ is the exciton translation mass, $k_{B}$ is the Boltzmann
constant, and $\hbar $ is the Planck constant, $i=s,p$, $\Gamma
_{0,s}(k_{\parallel })$ and $\Gamma _{0,p}(k_{\parallel })$~are radiative
decay rates of $s$- and $p$-polarised excitons.

\begin{figure}[hptb]
\includegraphics[width=0.47\textwidth]{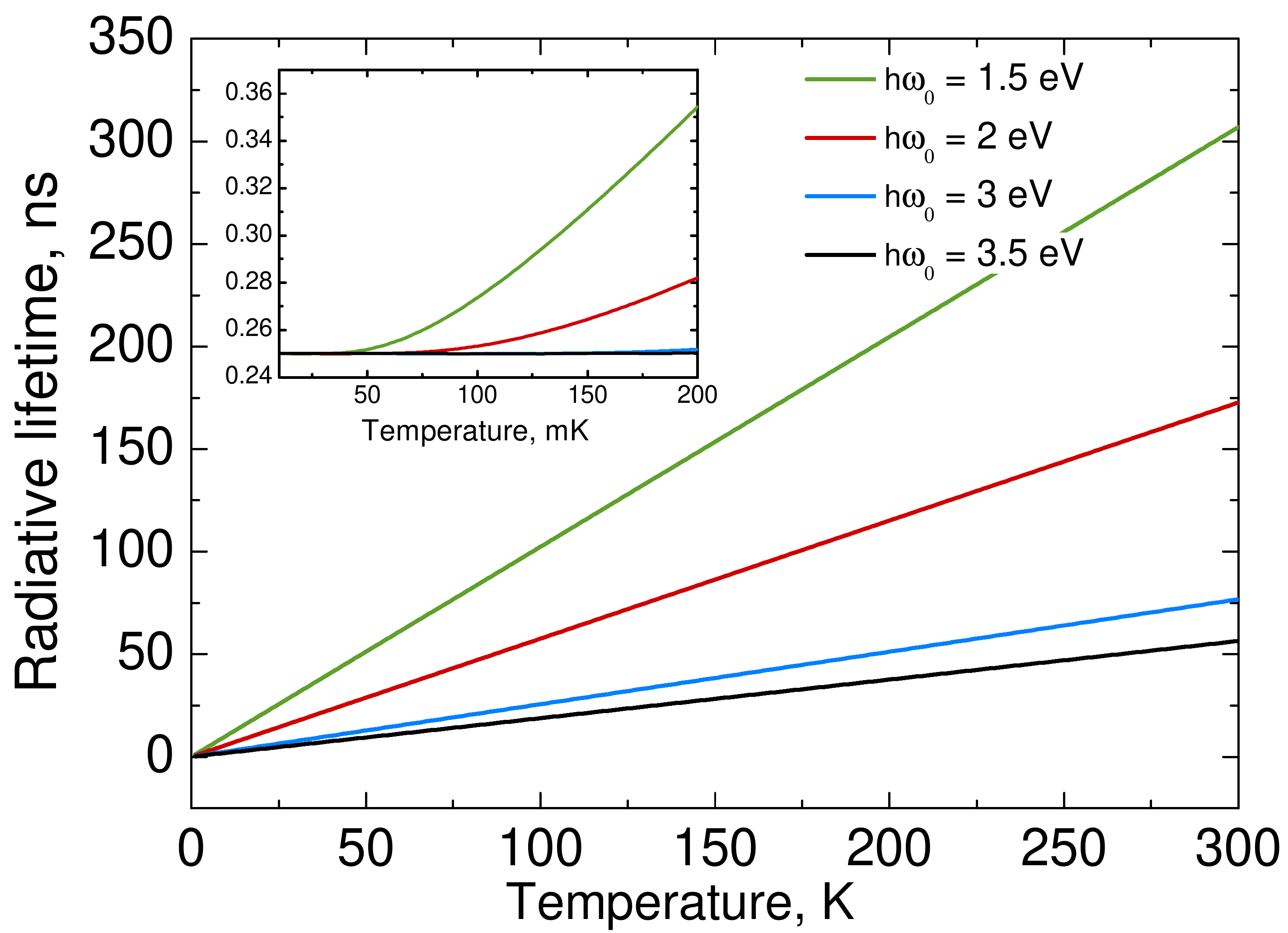}
\caption{Integrated exciton radiative lifetime for different temperature
regimes. The inset shows a nonlinear behaviour observed at low temperatures.
Different colours correspond to different values of exciton resonance
frequency. The parameters used $M = 0.1$, $n=2.47$ and $\protect\tau_0 =
1/(2\Gamma_0) = 1$~ns.}
\label{fig:fig2}
\end{figure}

Fig.~\ref{fig:fig2} shows the direct exciton radiative decay rate $%
1/(2\Gamma )$ as a function of temperature. It is seen that a linear
behaviour is observed in a range of temperatures where the typical kinetic
energy of exciton $\hbar ^{2}\omega ^{2}/(2Mc^{2})$ is much less than the
thermal energy $k_{B}T$. Nonlinear effects become apparent at
the temperatures of the order of tens of mK (see the inset).

\begin{figure}[hptb]
\includegraphics[width=0.43\textwidth]{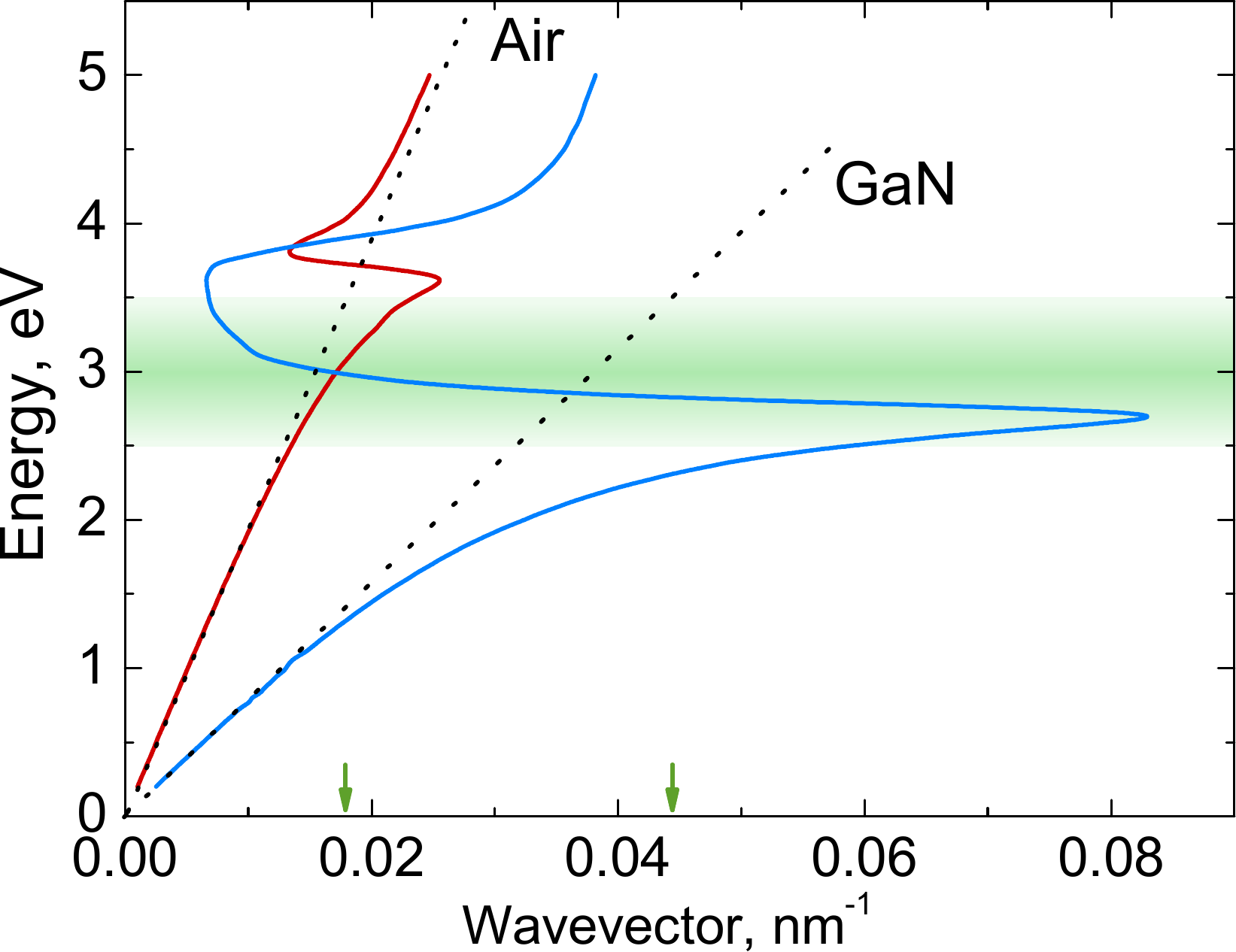}
\caption{The dispersion curves of the semi-infinite silver layer bound to
either air or GaN. Green arrows indicate the wavevector region $\protect%
\omega/c < k < n\protect\omega/c$.}
\label{fig:fig3}
\end{figure}

Surface plasmons propagate along a metal-dielectric interface with
wavevectors $\beta $ exceeding $\omega /c$ , which is why the amplitude of
electromagnetic field of a plasmon decays in both directions perpendicular
to the metal surface. The dispersion of these modes is given by~\cite%
{pitarke2007} 
\begin{equation}
\beta ^{2}=\frac{\varepsilon _{D}\varepsilon _{M}(\omega )}{\varepsilon
_{D}+\varepsilon _{M}(\omega )}\frac{\omega ^{2}}{c^{2}},
\label{eq:dispersion}
\end{equation}%
where $\varepsilon _{D,M}$ are dielectric functions of the dielectric
(semiconductor) and metal respectively. The frequency dependence of $%
\varepsilon _{M}$ in real metals is quite complicated. $\varepsilon _{M}$
has a large imaginary part accounting for the absorption of light in metal.
The useful characteristic of a plasmon is the extinction coefficient defined
as the imaginary part of effective dielectric function in the right part of
Eq.~(\ref{eq:dispersion}). The frequency dependence of the extinction
coefficient at the interfaces with air ($\varepsilon _{D}=1$) and GaN ($%
\varepsilon _{D}=6.1$~\cite{sanford:2980}) for aluminum, silver and gold
(experimental data on $\varepsilon _{M}(\omega )$ taken from~\cite%
{ehrenreich1962,ehrenreich1963,ehrenreich1965}) is presented in the lower
panel of Fig.~\ref{fig:fig1}. For the ease of comparison, the range of
exciton energies in InGaN with indium concentration varying from 0 to 0.3 is
shown in green. One can see that the most effective plasmon generation on
the both interfaces can be performed with the use of aluminum layer, but its
plasmon peak is quite far from the attractive range of frequencies. The gold
peak is close to that range, but the losses are too large to provide an
effective coupling, so we stop at the silver coating and use its parameters
for the following simulations. The dispersion curves of silver SPP's
calculated for the dielectric functions $\varepsilon _{M}(\omega )$
extracted from the experimental data are presented in Fig.~\ref{fig:fig3}.
The plasmon peak for silver approaches the exciton frequency in panel (a)
corresponding to the realistic GaN/metal interface.

The fundamental property of a smooth metal-semiconductor interface is that
neither the metal-semiconductors mode (SP1) nor the metal-air mode (SP2) can
be excited by the photons with $k_{\parallel }<\omega /c$. However, the
photons outside the light cone with $\omega /c<k<n\omega /c$, though unable
to escape the sample, may contribute to the excitation of the SP2 mode. This
leaky mode propagating on the metal-air interface can be excited as well due
to the exponential tail of its electric field, which penetrates into the
semiconductor region~\cite{maier_book, pitarke2007}. This additional decay
channel can be taken into account in a simple model of coupled oscillators~%
\cite{bellessa2008} with the coupling provided by the electric interaction
between exciton dipole momentum and the in-plane component of the plasmon
field. 
\begin{figure}[hptb]
\includegraphics[width=0.4\textwidth]{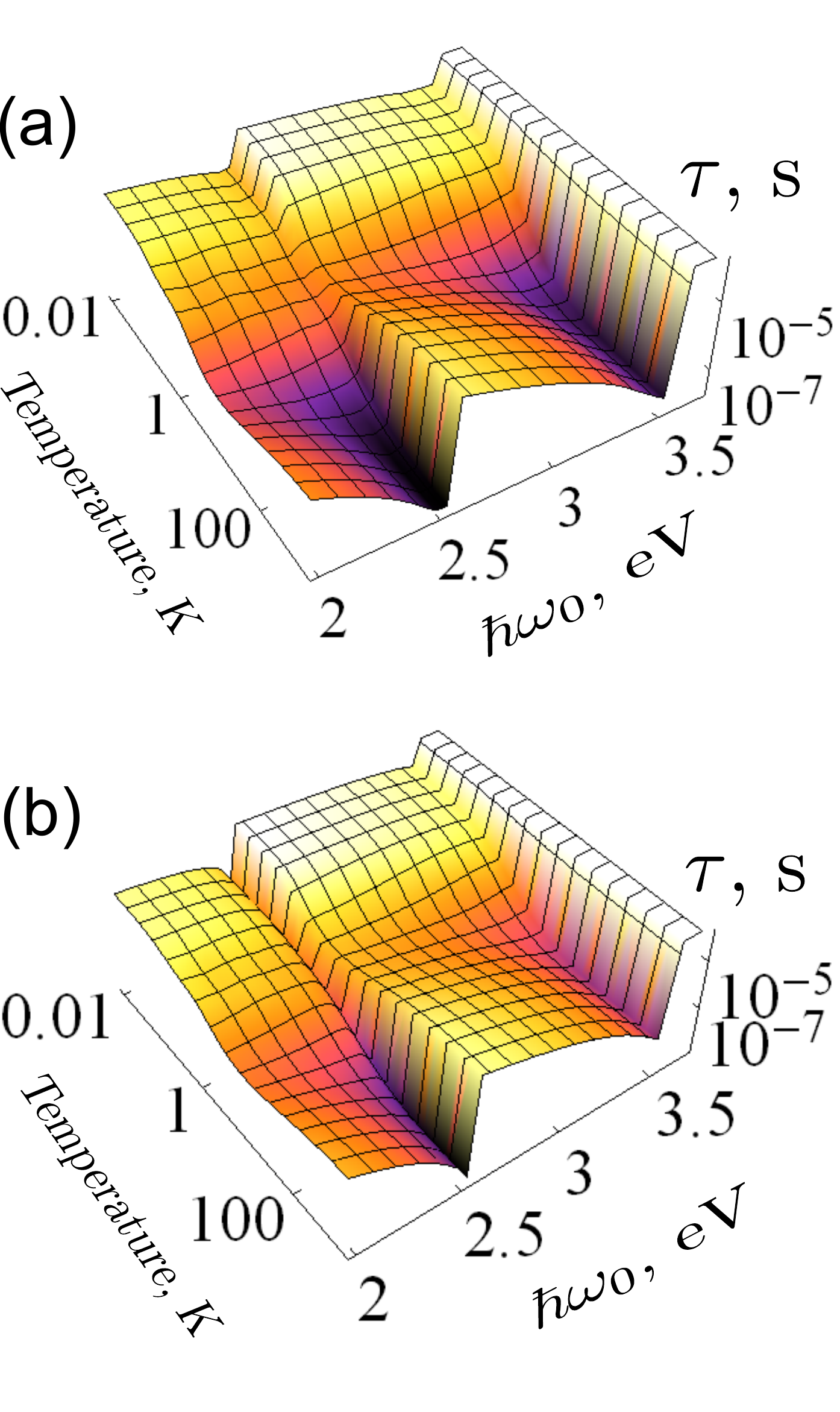}
\caption{Integrated exciton lifetime due to the leakage through SP2 and SP1
plasmon modes. The parameters are $\hbar \Gamma _{0}=3$~meV, $\protect \hbar \gamma =100$%
~meV, $a=d=5$~nm (a) $b=5$~nm and (b) $b=1$~nm.}
\label{fig:fig4}
\end{figure}
The detailed description of our model is given in Supplementary Information. Since the plasmon modes are TM-polarized they can be excited only by
p-polarized excitons with $\Gamma _{0,p}$ radiative decay. Exciton decay
through SP2 mode can be found from the following coupling equation 
\begin{equation}
\left[ \omega -\omega _{\mathrm{exc}}(k_{\parallel })\right] \left[ \omega
-\omega _{\mathrm{SP}}(k_{\parallel })+\mathrm{i}\gamma \right]
=V(k_{\parallel })^{2},  \label{eq:coupling}
\end{equation}%
where $\omega _{\mathrm{exc}}(k_{\parallel })$ and $\omega _{\mathrm{SP}%
}(k_{\parallel })$ are exciton and surface plasmon dispersions,
respectively, and $V(k_{\parallel })$ is the matrix element of coupling.
Here we have introduced the surface plasmon mode decay rate $\gamma $ with
two channels of decay: internal scattering in metal and scattering on the
roughness of the metal surface. The latter can be followed by an emission of
a photon, thus enhancing the light extraction from a sample. To emphasize
plasmonic effects we neglect here the exciton damping. The temperature
dependence of exciton decay through the plasmon mode can now be obtained by
taking an imaginary part of the exciton-like branch of Eq.~(\ref{eq:coupling}%
) and averaging it over the thermal distribution similarly to the Eq.~(\ref%
{eq:average}). If the surface of metal is rough or consists of clusters,
photons scattered at the interface may provide an additional channel for
radiative decay of the SP1 mode. We will further present the results of
calculations where both channels are taken into account. Fig.~\ref{fig:fig4}
shows the dependence of the exciton decay time through the plasmon modes on
temperature and exciton resonance energy $\hbar \omega _{0}$. The
thicknesses of metal, barrier and quantum well layers are denoted as $d$, $b$
and $a$, respectively. Note that the frequency dependence of the decay time
exhibits two dips at the positions of plasmonic resonances. As for the
temperature behavior, it is strongly nonmonotonous and in the frequency
region where both plasmon resonances are present ($\omega _{0}\lesssim 2.56$%
~eV for silver) it also exhibits two minima. Those minima become apparent if
the thermal distribution function has its maximum in the region of
anti-crossing between exciton and plasmon branches. Once the exciton
resonant frequency moves to higher values one of the minima vanishes since
only one plasmon mode is generated in this regime. 
\begin{figure}[hptb]
\includegraphics[width=0.47\textwidth]{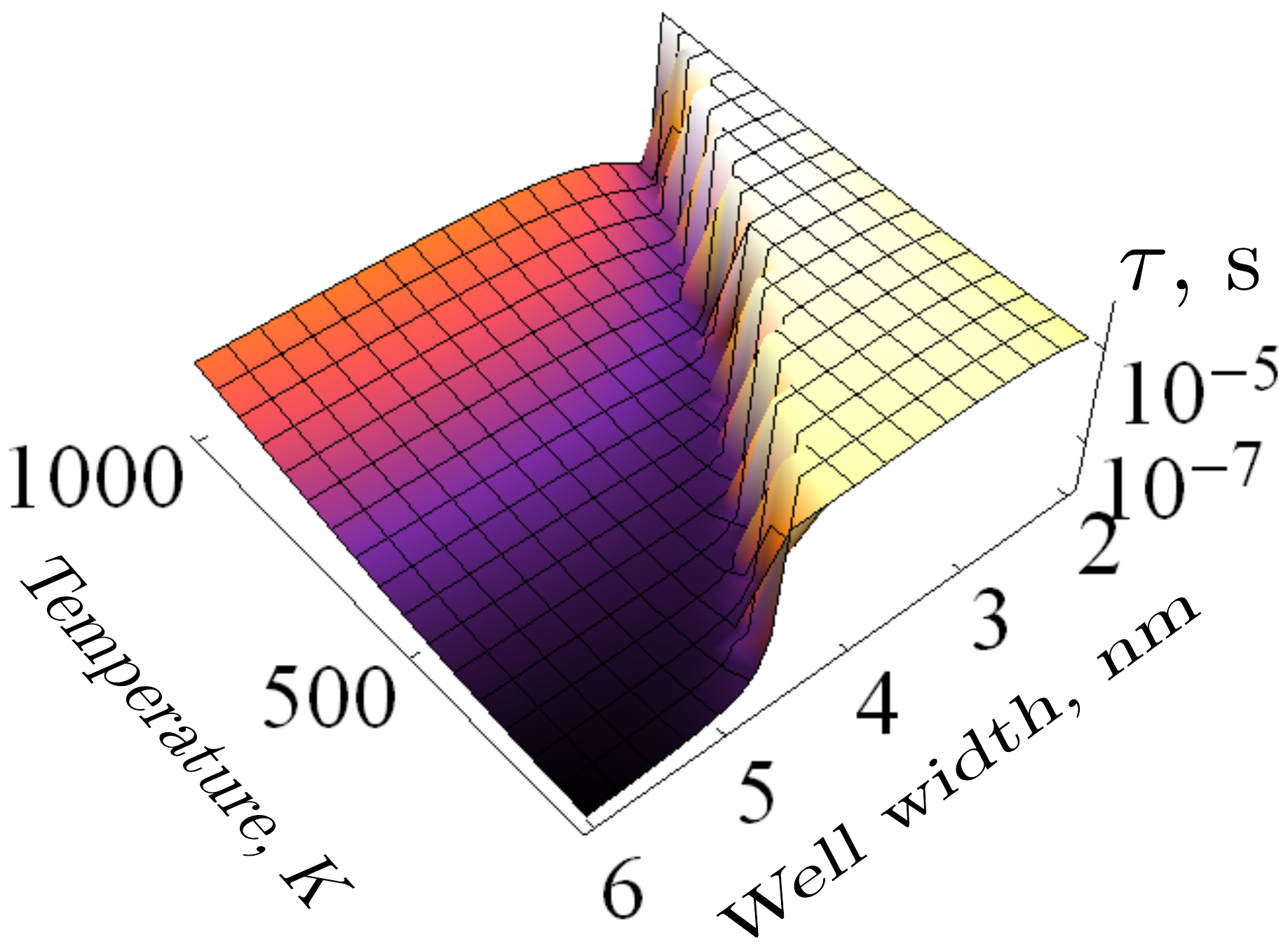}
\caption{Integrated exciton lifetime versus temperature and well width.}
\label{fig:fig5}
\end{figure}

The exciton energy in a quantum well can be tuned to the plasmon resonance
by variation of the well width and temperature. This tuning results in a
significant decrease of the exciton lifetime, which is illustrated by Fig.~%
\ref{fig:fig5}. This figure offer a tool for life-time engineering in
metal-semiconductor structures: tuning the thicknesses of well and barrier
layers one can vary the exciton radiative lifetime by orders of magnitude.

In conclusion, we have shown that the coupling of excitons in quantum wells
with the surface plasmons at metal-semiconductor and metal-air interfaces
yileds additional channels of exciton radiative decay which may
significantly reduce the exciton radiative lifetime. The temperature
dependence of exciton radiative life-time becomes strongly non-monotonic in
the presence of plasmons. It experiences two dips as the exciton gas heats up
to the energies of two plasmon modes in an air-metal-semiconductor structure. The plasmon-controlled Purcell effect opens way to the life-time engineering in semiconductor-metal quantum confined structures which may be widely used in solid state light emitters.


\begin{thebibliography}
 \expandafter\ifx\csname natexlab\endcsname\relax

\fi
\expandafter\ifx\csname bibnamefont\endcsname\relax

\fi
\expandafter\ifx\csname bibfnamefont\endcsname\relax

\fi
\expandafter\ifx\csname citenamefont\endcsname\relax

\fi
\expandafter\ifx\csname url\endcsname\relax

\fi
\expandafter\ifx\csname urlprefix\endcsname\relax

\fi
\providecommand{\bibinfo}[2]{#2} \providecommand{\eprint}[2][]{\url{#2}}

\bibitem[Koenderink(2010)]{konderink} \bibinfo{author}{\bibfnamefont{A.~F.}
\bibnamefont{Koenderink}},  \bibinfo{journal}{Optics Letters} \textbf{%
\bibinfo{volume}{35}},  \bibinfo{pages}{4208} (\bibinfo{year}{2010}).

\bibitem[Meinzer et~al.(2010)Meinzer, Ruther, Linden, Soukoulis, Khitrova,
Hendrickson, Olitzky, Gibbs, and Wegener]{meinzer} \bibinfo{author}{%
\bibfnamefont{N.}~\bibnamefont{Meinzer}},  \bibinfo{author}{%
\bibfnamefont{M.}~\bibnamefont{Ruther}},  \bibinfo{author}{%
\bibfnamefont{S.}~\bibnamefont{Linden}},  \bibinfo{author}{%
\bibfnamefont{C.~M.} \bibnamefont{Soukoulis}},  \bibinfo{author}{%
\bibfnamefont{G.}~\bibnamefont{Khitrova}},  \bibinfo{author}{%
\bibfnamefont{J.}~\bibnamefont{Hendrickson}},  \bibinfo{author}{%
\bibfnamefont{J.~D.} \bibnamefont{Olitzky}},  \bibinfo{author}{%
\bibfnamefont{H.~M.} \bibnamefont{Gibbs}}, and  \bibinfo{author}{%
\bibfnamefont{M.}~\bibnamefont{Wegener}},  \bibinfo{journal}{Opt. Express 18}
\textbf{\bibinfo{volume}{18}},  \bibinfo{pages}{24140} (\bibinfo{year}{2010}%
).

\bibitem[Anger et~al.(2006)Anger, Bharadwaj, and Novotny]%
{PhysRevLett.96.113002} \bibinfo{author}{\bibfnamefont{P.}~%
\bibnamefont{Anger}},  \bibinfo{author}{\bibfnamefont{P.}~%
\bibnamefont{Bharadwaj}},  and \bibinfo{author}{\bibfnamefont{L.}~%
\bibnamefont{Novotny}},  \bibinfo{journal}{Phys. Rev. Lett.} \textbf{%
\bibinfo{volume}{96}},  \bibinfo{pages}{113002} (\bibinfo{year}{2006}).

\bibitem[K\"uhn et~al.(2006)K\"uhn, H\aa {}kanson, Rogobete, and Sandoghdar]%
{PhysRevLett.97.017402} \bibinfo{author}{\bibfnamefont{S.}~\bibnamefont{K%
\"uhn}},  \bibinfo{author}{\bibfnamefont{U.}~\bibnamefont{H\aa{}kanson}},  %
\bibinfo{author}{\bibfnamefont{L.}~\bibnamefont{Rogobete}}, and  %
\bibinfo{author}{\bibfnamefont{V.}~\bibnamefont{Sandoghdar}},  %
\bibinfo{journal}{Phys. Rev. Lett.} \textbf{\bibinfo{volume}{97}},  %
\bibinfo{pages}{017402} (\bibinfo{year}{2006}).

\bibitem[Curto et~al.(2010)Curto, Volpe, Taminiau, Kreuzer, Quidant, and van
Hulst]{science_curto} \bibinfo{author}{\bibfnamefont{A.~G.}
\bibnamefont{Curto}},  \bibinfo{author}{\bibfnamefont{G.}~%
\bibnamefont{Volpe}},  \bibinfo{author}{\bibfnamefont{T.~H.}
\bibnamefont{Taminiau}},  \bibinfo{author}{\bibfnamefont{M.~P.}
\bibnamefont{Kreuzer}},  \bibinfo{author}{\bibfnamefont{R.}~%
\bibnamefont{Quidant}}, and  \bibinfo{author}{\bibfnamefont{N.~F.}
\bibnamefont{van Hulst}},  \bibinfo{journal}{Science} \textbf{%
\bibinfo{volume}{329}},  \bibinfo{pages}{930} (\bibinfo{year}{2010}).

\bibitem[Ajiki et~al.(2002)Ajiki, Tsuji, Kawano, and Cho]{Ajiki} %
\bibinfo{author}{\bibfnamefont{H.}~\bibnamefont{Ajiki}},  %
\bibinfo{author}{\bibfnamefont{T.}~\bibnamefont{Tsuji}},  %
\bibinfo{author}{\bibfnamefont{K.}~\bibnamefont{Kawano}}, and  %
\bibinfo{author}{\bibfnamefont{K.}~\bibnamefont{Cho}},  %
\bibinfo{journal}{Phys. Rev. B} \textbf{\bibinfo{volume}{66}},  %
\bibinfo{pages}{245322} (\bibinfo{year}{2002}).

\bibitem[Glazov et~al.(2011)Glazov, Ivchenko, Poddubny, and Khitrova]%
{glazov_ftt} \bibinfo{author}{\bibfnamefont{M.~M.} \bibnamefont{Glazov}},  %
\bibinfo{author}{\bibfnamefont{E.~L.} \bibnamefont{Ivchenko}},  %
\bibinfo{author}{\bibfnamefont{A.~N.} \bibnamefont{Poddubny}},  and %
\bibinfo{author}{\bibfnamefont{G.}~\bibnamefont{Khitrova}},  %
\bibinfo{journal}{Fiz. Tverd. Tela} \textbf{\bibinfo{volume}{53}},  %
\bibinfo{pages}{1665} (\bibinfo{year}{2011}).

\bibitem[Ford and Weber(1984)]{Ford1984} \bibinfo{author}{%
\bibfnamefont{G.~W.} \bibnamefont{Ford}} and  \bibinfo{author}{%
\bibfnamefont{W.~H.} \bibnamefont{Weber}},  \bibinfo{journal}{Phys. Rep.} 
\textbf{\bibinfo{volume}{113}},  \bibinfo{pages}{195} (\bibinfo{year}{1984}).

\bibitem[Chance et~al.(1978)Chance, Prock, and Silbey]{Chance1978} %
\bibinfo{author}{\bibfnamefont{R.~R.} \bibnamefont{Chance}},  %
\bibinfo{author}{\bibfnamefont{A.}~\bibnamefont{Prock}}, and  %
\bibinfo{author}{\bibfnamefont{R.}~\bibnamefont{Silbey}},  %
\bibinfo{journal}{Adv. Chem. Phys.} \textbf{\bibinfo{volume}{37}},  %
\bibinfo{pages}{1} (\bibinfo{year}{1978}).

\bibitem[Gontijo et~al.(1999)Gontijo, Boroditsky, Yablonovitch, Keller,
Mishra, and DenBaars]{yablonovitch1999} \bibinfo{author}{\bibfnamefont{I.}~%
\bibnamefont{Gontijo}},  \bibinfo{author}{\bibfnamefont{M.}~%
\bibnamefont{Boroditsky}},  \bibinfo{author}{\bibfnamefont{E.}~%
\bibnamefont{Yablonovitch}},  \bibinfo{author}{\bibfnamefont{S.}~%
\bibnamefont{Keller}},  \bibinfo{author}{\bibfnamefont{U.~K.}
\bibnamefont{Mishra}},  and 
\bibinfo{author}{\bibfnamefont{S.~P.}
  \bibnamefont{DenBaars}}, \bibinfo{journal}{Phys. Rev. B}  \textbf{%
\bibinfo{volume}{60}}, \bibinfo{pages}{11564} (\bibinfo{year}{1999}).

\bibitem[Neogi et~al.(2002)Neogi, Lee, Everitt, Kuroda, Tackeuchi, and
Yablonovitch]{yablonovitch2002} \bibinfo{author}{\bibfnamefont{A.}~%
\bibnamefont{Neogi}},  \bibinfo{author}{\bibfnamefont{C.-W.}
\bibnamefont{Lee}},  \bibinfo{author}{\bibfnamefont{H.~O.}
\bibnamefont{Everitt}},  \bibinfo{author}{\bibfnamefont{T.}~%
\bibnamefont{Kuroda}},  \bibinfo{author}{\bibfnamefont{A.}~%
\bibnamefont{Tackeuchi}},  and  \bibinfo{author}{\bibfnamefont{E.}~%
\bibnamefont{Yablonovitch}},  \bibinfo{journal}{Phys. Rev. B} \textbf{%
\bibinfo{volume}{66}},  \bibinfo{pages}{153305} (\bibinfo{year}{2002}).

\bibitem[Okamoto et~al.(2004)Okamoto, Niki, Shvartser, Narukawa, Mukai, and
Scherer]{nat_mat_okamoto} \bibinfo{author}{\bibfnamefont{K.}~%
\bibnamefont{Okamoto}},  \bibinfo{author}{\bibfnamefont{I.}~%
\bibnamefont{Niki}},  \bibinfo{author}{\bibfnamefont{A.}~%
\bibnamefont{Shvartser}},  \bibinfo{author}{\bibfnamefont{Y.}~%
\bibnamefont{Narukawa}},  \bibinfo{author}{\bibfnamefont{T.}~%
\bibnamefont{Mukai}}, and  \bibinfo{author}{\bibfnamefont{A.}~%
\bibnamefont{Scherer}},  \bibinfo{journal}{Nat. Mater.} \textbf{%
\bibinfo{volume}{3}},  \bibinfo{pages}{601 } (\bibinfo{year}{2004}).

\bibitem[Okamoto et~al.(2005)Okamoto, Niki, Scherer, Narukawa, Mukai, and
Kawakami]{okamoto:071102} \bibinfo{author}{\bibfnamefont{K.}~%
\bibnamefont{Okamoto}},  \bibinfo{author}{\bibfnamefont{I.}~%
\bibnamefont{Niki}},  \bibinfo{author}{\bibfnamefont{A.}~%
\bibnamefont{Scherer}},  \bibinfo{author}{\bibfnamefont{Y.}~%
\bibnamefont{Narukawa}},  \bibinfo{author}{\bibfnamefont{T.}~%
\bibnamefont{Mukai}}, and  \bibinfo{author}{\bibfnamefont{Y.}~%
\bibnamefont{Kawakami}},  \bibinfo{journal}{Applied Physics Letters} \textbf{%
\bibinfo{volume}{87}},  \bibinfo{eid}{071102} (pages~\bibinfo{numpages}{3}) (%
\bibinfo{year}{2005}).

\bibitem[Lin et~al.(2010)Lin, Mohammadizia, Neogi, Morkoc, and Ohtsu]%
{lin:221104} \bibinfo{author}{\bibfnamefont{J.}~\bibnamefont{Lin}},  %
\bibinfo{author}{\bibfnamefont{A.}~\bibnamefont{Mohammadizia}},  %
\bibinfo{author}{\bibfnamefont{A.}~\bibnamefont{Neogi}},  %
\bibinfo{author}{\bibfnamefont{H.}~\bibnamefont{Morkoc}}, and  %
\bibinfo{author}{\bibfnamefont{M.}~\bibnamefont{Ohtsu}},  %
\bibinfo{journal}{Applied Physics Letters} \textbf{\bibinfo{volume}{97}},  %
\bibinfo{eid}{221104} (pages~\bibinfo{numpages}{3}) (\bibinfo{year}{2010}).

\bibitem[Jang et~al.(2012)Jang, Jeon, Sahoo, Jo, Ju, jae Lee, Baek, Yang,
Song, Polyakov et~al.]{Jang:12} \bibinfo{author}{\bibfnamefont{L.-W.}
\bibnamefont{Jang}},  \bibinfo{author}{\bibfnamefont{D.-W.}
\bibnamefont{Jeon}},  \bibinfo{author}{\bibfnamefont{T.}~\bibnamefont{Sahoo}}%
,  \bibinfo{author}{\bibfnamefont{D.-S.} \bibnamefont{Jo}},  %
\bibinfo{author}{\bibfnamefont{J.-W.} \bibnamefont{Ju}},  %
\bibinfo{author}{\bibfnamefont{S.}~\bibnamefont{jae Lee}},  %
\bibinfo{author}{\bibfnamefont{J.-H.} \bibnamefont{Baek}},  %
\bibinfo{author}{\bibfnamefont{J.-K.} \bibnamefont{Yang}},  %
\bibinfo{author}{\bibfnamefont{J.-H.} \bibnamefont{Song}},  %
\bibinfo{author}{\bibfnamefont{A.~Y.} \bibnamefont{Polyakov}},  et~al., %
\bibinfo{journal}{Opt. Express}  \textbf{\bibinfo{volume}{20}}, %
\bibinfo{pages}{2116} (\bibinfo{year}{2012}).

\bibitem[Toropov et~al.(2009)Toropov, Shubina, Jmerik, Ivanov, Ogawa, and
Minami]{toropovPRL2009} \bibinfo{author}{\bibfnamefont{A.~A.}
\bibnamefont{Toropov}},  \bibinfo{author}{\bibfnamefont{T.~V.}
\bibnamefont{Shubina}},  \bibinfo{author}{\bibfnamefont{V.~N.}
\bibnamefont{Jmerik}},  \bibinfo{author}{\bibfnamefont{S.~V.}
\bibnamefont{Ivanov}},  \bibinfo{author}{\bibfnamefont{Y.}~%
\bibnamefont{Ogawa}}, and  \bibinfo{author}{\bibfnamefont{F.}~%
\bibnamefont{Minami}},  \bibinfo{journal}{Phys. Rev. Lett.} \textbf{%
\bibinfo{volume}{103}},  \bibinfo{pages}{037403} (\bibinfo{year}{2009}).

\bibitem[Tanaka et~al.(2010)Tanaka, Plum, Ou, Uchino, and Zheludev]%
{TanakaPRL2010} \bibinfo{author}{\bibfnamefont{K.}~\bibnamefont{Tanaka}},  %
\bibinfo{author}{\bibfnamefont{E.}~\bibnamefont{Plum}},  \bibinfo{author}{%
\bibfnamefont{J.~Y.} \bibnamefont{Ou}},  \bibinfo{author}{\bibfnamefont{T.}~%
\bibnamefont{Uchino}}, and  \bibinfo{author}{\bibfnamefont{N.~I.}
\bibnamefont{Zheludev}},  \bibinfo{journal}{Phys. Rev. Lett.} \textbf{%
\bibinfo{volume}{105}},  \bibinfo{pages}{227403} (\bibinfo{year}{2010}).

\bibitem[Ivchenko(2005)]{ivchenko05a} \bibinfo{author}{\bibfnamefont{E.~L.}
\bibnamefont{Ivchenko}},  \emph{\bibinfo{title}{Optical spectroscopy of
semiconductor nanostructures}}  (\bibinfo{publisher}{Alpha Science, Harrow
UK}, \bibinfo{year}{2005}).

\bibitem[Pitarke et~al.(2007)Pitarke, Silkin, Chulkov, and Echenique]%
{pitarke2007} \bibinfo{author}{\bibfnamefont{J.~M.} \bibnamefont{Pitarke}},  %
\bibinfo{author}{\bibfnamefont{V.~M.} \bibnamefont{Silkin}},  %
\bibinfo{author}{\bibfnamefont{E.~V.} \bibnamefont{Chulkov}},  and 
\bibinfo{author}{\bibfnamefont{P.~M.}
  \bibnamefont{Echenique}}, \bibinfo{journal}{Rep. Prog. Phys.}  \textbf{%
\bibinfo{volume}{70}}, \bibinfo{pages}{1} (\bibinfo{year}{2007}).

\bibitem[Sanford et~al.(2003)Sanford, Robins, Davydov, Shapiro, Tsvetkov,
Dmitriev, Keller, Mishra, and DenBaars]{sanford:2980} \bibinfo{author}{%
\bibfnamefont{N.~A.} \bibnamefont{Sanford}},  \bibinfo{author}{%
\bibfnamefont{L.~H.} \bibnamefont{Robins}},  \bibinfo{author}{%
\bibfnamefont{A.~V.} \bibnamefont{Davydov}},  \bibinfo{author}{%
\bibfnamefont{A.}~\bibnamefont{Shapiro}},  \bibinfo{author}{%
\bibfnamefont{D.~V.} \bibnamefont{Tsvetkov}},  \bibinfo{author}{%
\bibfnamefont{A.~V.} \bibnamefont{Dmitriev}},  \bibinfo{author}{%
\bibfnamefont{S.}~\bibnamefont{Keller}},  \bibinfo{author}{%
\bibfnamefont{U.~K.} \bibnamefont{Mishra}},  and 
\bibinfo{author}{\bibfnamefont{S.~P.}
  \bibnamefont{DenBaars}}, \bibinfo{journal}{Journal of Applied Physics}  
\textbf{\bibinfo{volume}{94}}, \bibinfo{pages}{2980} (\bibinfo{year}{2003}).

\bibitem[Ehrenreich and Philipp(1962)]{ehrenreich1962} \bibinfo{author}{%
\bibfnamefont{H.}~\bibnamefont{Ehrenreich}} and  \bibinfo{author}{%
\bibfnamefont{H.}~\bibnamefont{Philipp}},  \bibinfo{journal}{Physical Review}
\textbf{\bibinfo{volume}{128}},  \bibinfo{pages}{1622} (\bibinfo{year}{1962}%
).

\bibitem[Ehrenreich et~al.(1963)Ehrenreich, Philipp, and Segall]%
{ehrenreich1963} \bibinfo{author}{\bibfnamefont{H.}~\bibnamefont{Ehrenreich}}%
,  \bibinfo{author}{\bibfnamefont{H.}~\bibnamefont{Philipp}}, and  %
\bibinfo{author}{\bibfnamefont{B.}~\bibnamefont{Segall}},  %
\bibinfo{journal}{Physical Review} \textbf{\bibinfo{volume}{132}},  %
\bibinfo{pages}{1918} (\bibinfo{year}{1963}).

\bibitem[Cooper et~al.(1965)Cooper, Ehrenreich, and Philipp]{ehrenreich1965} %
\bibinfo{author}{\bibfnamefont{B.~R.} \bibnamefont{Cooper}},  %
\bibinfo{author}{\bibfnamefont{H.}~\bibnamefont{Ehrenreich}},  and 
\bibinfo{author}{\bibfnamefont{H.~R.}
  \bibnamefont{Philipp}}, \bibinfo{journal}{Physical Review}  \textbf{%
\bibinfo{volume}{138}}, \bibinfo{pages}{A494} (\bibinfo{year}{1965}).

\bibitem[Maier(2007)]{maier_book} \bibinfo{author}{\bibfnamefont{S.}~%
\bibnamefont{Maier}},  \emph{\bibinfo{title}{Plasmoncs. Fundamentals and
Applications}}  (\bibinfo{publisher}{Springer}, \bibinfo{year}{2007}).

\bibitem[Bellessa et~al.(2008)Bellessa, Symonds, Meynaud, Plenet, Cambril,
Miard, Ferlazzo, and Lemaitre]{bellessa2008} \bibinfo{author}{%
\bibfnamefont{J.}~\bibnamefont{Bellessa}},  \bibinfo{author}{%
\bibfnamefont{C.}~\bibnamefont{Symonds}},  \bibinfo{author}{%
\bibfnamefont{C.}~\bibnamefont{Meynaud}},  \bibinfo{author}{%
\bibfnamefont{J.~C.} \bibnamefont{Plenet}},  \bibinfo{author}{%
\bibfnamefont{E.}~\bibnamefont{Cambril}},  \bibinfo{author}{%
\bibfnamefont{A.}~\bibnamefont{Miard}},  \bibinfo{author}{\bibfnamefont{L.}~%
\bibnamefont{Ferlazzo}}, and  \bibinfo{author}{\bibfnamefont{A.}~%
\bibnamefont{Lemaitre}},  \bibinfo{journal}{Phys. Rev. B} \textbf{%
\bibinfo{volume}{78}},  \bibinfo{pages}{205326} (\bibinfo{year}{2008}).
\end{thebibliography}
\end{document}